# Playing Billiard in Version Space

Pál Ruján[*]

*Fachbereich 8 Physik and ICBM, Postfach 2503*

*Carl-von-Ossietzky Universität*

*26111 Oldenburg, Germany*

## Abstract

A ray-tracing method inspired by ergodic billiards is used to estimate the theoretically best decision rule for a set of linear separable examples. While the Bayes-optimum requires in general a majority decision over *all* Perceptrons separating the example set, the problem considered here corresponds to finding the *single* Perceptron with best average generalization probability. For randomly distributed examples the billiard estimate agrees with known analytic results, while for real-life classification problems it reduces consistently the generalization error compared to the maximal stability Perceptron.

PACS numbers: 87.10+e, 02.50-r, 05.90+m

## 1 Introduction

The problem of supervised learning for the class of linearly separable functions corresponds to the Perceptron problem [1] and its variants. Given a set of $M$ examples $\{\vec{\xi}^{(\alpha)}\}_{\alpha=1}^{M}$, $\vec{\xi}^{(\alpha)} \in \mathbb{R}^N$ and their corresponding binary class label $\sigma_\alpha = \pm 1$ generated independently and randomly from the distribution $P(\vec{\xi}, \sigma)$ one is looking for the hyperplane $(\vec{w}, \vec{x}) = \theta$ separating the positive from the negative labeled examples

---

[*]e-mail:rujan@neuro.uni-oldenburg.de

$$\begin{aligned}
(\vec{w}, \vec{\xi}^{(\alpha)}) &\geq \theta_1, \; \sigma_\alpha = 1 \\
(\vec{w}, \vec{\xi}^{(\beta)}) &\leq \theta_2, \; \sigma_\beta = -1 \\
\theta_1 &> \theta > \theta_2
\end{aligned} \quad (1)$$

subject to some optimality criterion. Very often the task is to minimize the average probability of class error of a $\vec{x}$ vector sampled randomly from $P(\vec{\xi}, \sigma)$. Other optimization criteria are obtained by minimizing the maximal error (worst case optimization), by requiring a minimal average classification error for noisy examples, etc. In what follows we restrict ourselves to the 'standard' problem of minimal average classification error for (noiseless) examples. The vector space whose points are the vectors $(\vec{w}, \theta)$ is called the 'version space' and its subspace satisfying Eq. (1) the solution polyhedron.

The problem has both theoretical and practical aspects. In theory, one knows that the optimal Bayes approach [2] implies an average over all $\{\vec{w}, \theta\}$ solutions satisfying Eq. (1). In presence of noise, this average should be taken over a Boltzmann-type distribution at an appropriate effective temperature [3]. For randomly distributed examples ($P(\vec{\xi}, \sigma) = const$) Watkin [4] has shown that the average over all possible noiseless solutions is equivalent to computing the center of mass of the solution polyhedron. On the practical side, known learning algorithms like Adaline [5] or the maximal stability Perceptron (MSP) [6, 7, 8] are not Bayes-optimal. In fact, as shown below, the maximal stability Perceptron corresponds to the center of the largest hypersphere inscribed into the solution polyhedron.

There have been several attempts at developing learning algorithms approximating the Bayes-optimum. Early on, Monte Carlo simulations have been used for sampling the version space [4]. More recent methods aim at estimating the center of mass of the solution space. Bouten *et al* use an appropriately defined convex function whose unique minimum is tuned with the help of a single parameter as to approximate the Bayes solution [9]. A somewhat similar notion of an 'analytic center of a convex polytope' introduced by Sonnevend [10] has been used extensively for designing fast linear programming algorithms [11]. This algorithm has been tested on *randomly distributed* examples generated from a teacher Perceptron, where analytical results can be used to determine the correct parameter value. The question whether the method can be also used in other instances is still open. A very promising but not yet fully exploited approach [12] uses the Thouless-Anderson-Palmer (TAP) type equations. In this paper we present a rather different approach, based on an analogy to classical ergodic billiards. Considering the solution polyhedron as a dynamic system, a long trajectory is generated and used to estimate the center of mass of the billiard.

Questions related to ergodic theory of billiards are briefly considered in Section 2. Section 3 sets up the stage by presenting an elementary geometric analysis in two dimensions. The underlying version-space problem is discussed for the Perceptron in Section 4. An elementary algorithmic implementation for open polyhedral cones and for their projective geometric closures are summarized in Section 5. Numerical results and a comparison to known analytic bounds and to other learning algorithms can be found in Section 6. Our conclusions and further prospects are summarized in Section 7.

## 2 Billiards

A *billiard* is usually defined as a closed space region (compact set) $\mathcal{P} \in \mathbb{R}^N$ dimensions. The boundaries of a billiard are usually piecewise smooth functions. The dynamics is the free motion of a point mass (ball) undergoing elastic collisions with the enclosing walls. Hence, the absolute value of the momentum is preserved and the phase space is the direct product of $\mathcal{B} = \mathcal{P} \times \mathcal{S}^{N-1}$, where $\mathcal{S}^{N-1}$ is the surface of the $N$-dimensional unit velocity sphere. Such a simple Hamiltonian dynamics defines a flow and its Poincaré map an automorphism. The mathematicians have defined a finely tuned hierarchy of notions related to such dynamic systems. For instance, simple ergodicity as implied by the Birkhoff-Hincsin theorem means that the average of any integrable function defined on the phase space over a single but very long trajectory equals the spatial mean (except for a set of zero measure). Furthermore, integrable functions invariant under the dynamics must be constant. From a practical point of view this means that almost all infinitely long trajectories cover uniformly the phase space. Properties like mixing (Kolmogorov-mixing) are stronger than ergodicity and require that the flow will eventually fully (and uniformly) mix different subsets of $\mathcal{B}$. In hyperbolic systems one can go further and construct Markov partitions defined on symbolic dynamics and eventually prove related central limit the-orems.

Not all convex billiards are ergodic. Notable exceptions are ellipsoidal billiards, which can be solved by a proper separation of variables [13]. Already Jacobi knew that a trajectory started close to and along the boundaries of an ellipse cannot reach a central region bounded by the so-called caustics. In addition to separable billiards there are a few other exactly soluble polyhedral billiards [14]. Such solutions are intimately related to the reflection method - the billiard tiles perfectly the entire space. Examples of integrable but non separable billiards can be obtained by mapping exactly soluble low dimensional many particle system into a one-particle high-dimensional billiard [15].

On the other hand, the stadium billiard (two half-circles joined by two parallel lines) is ergodic in a strong sense, the metric entropy is non-vanishing [16]. The dynamics induced by the billiard is hyperbolic if at any point in phase space there are both expanding (unstable) and shrinking (stable) manifolds. A famous example is Sinai's reformulation of the Lorentz-gas [17]. Deep mathematical methods were needed to prove the Kolmogorov-mixing property and in constructing the Markov partitions for the symbolic dynamics [18].

The question whether a particular billiard is ergodic or not can be decided in principle by solving the Schrödinger problem for a free particles moving in the billiard-box. If the eigenfunctions corresponding to the high energy modes are roughly constant, then the billiard is ergodic. Only few results are known for such quantum billiards.

I am not aware of a general theory regarding the ergodic properties of convex polyhedral billiards. If all angles of the polyhedra are rational, then the billiard is weakly ergodic in the sense that the velocity direction will reach only rational angles (relative to the initial direction). In general, as long as two neighboring trajectories collide with the same polyhedral faces, their distance will grow only linearly. Once they are far enough as to collide with different faces of the polyhedra, their distance will abruptly increase. Except for very special cases with high symmetry, it seems therefore unlikely that high dimensional convex polyhedra are not ergodic. If so, adding a few scatterers *inside* the polyhedra might restore ergodicity.

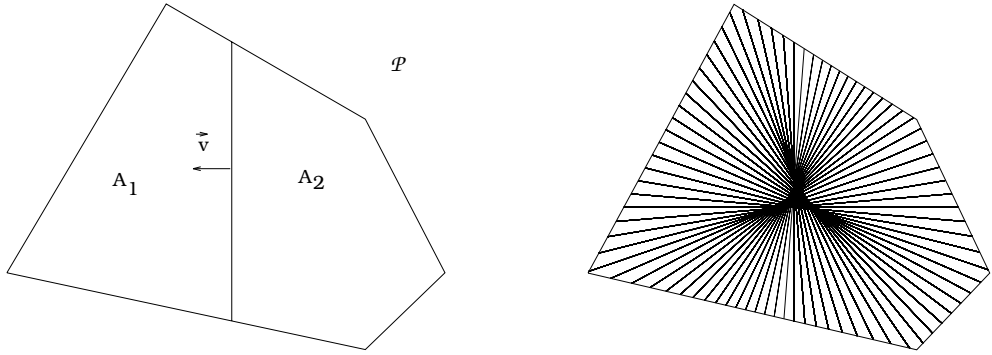

Figure 1: *a) Halving the surface along a given direction and b) the resulting Bayes-lines*

## 3 A simple geometric problem

For the sake of simplicity let us illustrate our approach in a two-dimensional setting. Although the problem considered below is interesting in its own, the geometry of the Perceptron version space polyhedron is slightly different (see the next Section for details).

Let $\mathcal{P}$ be a closed convex polygon and $\vec{v}$ a given unit vector, defining a particular direction in the $\mathbb{R}^2$ space. Next, construct the line perpendicular to $\vec{v}$ which halves the surface of the polygon $\mathcal{P}$, $A_1 = A_2$, as illustrated in Fig. 1a. Repeating this construction for a set of equally spaced angles leads to the 'Bayes-lines' seen in Fig. 1b. Given a particular normal vector, the Bayes decision rule requires the computation of the two volumes $A_1$ and $A_2$ in a high dimensional space, a very demanding task. From a practical point of view it is certainly more economical to compute and store a single point $\vec{r}_0$, which represents 'optimally' the full information contained in Fig. 1b.

The lines passing through $\vec{r}_0$ do not partition $\mathcal{P}$ in equal parts but make some mistakes, $\Delta A = A_1 - A_2 \neq 0$ which depend on both $\vec{v}$ and the polygon $\mathcal{P}$, $\Delta A = \Delta A(\vec{v}, \mathcal{P})$. Different optimality criteria can be formulated, depending on the actual application. The usual approach corresponds to minimizing the average squared area-difference:

$$\vec{r}_0 = \arg\min \langle (\Delta A)^2 \rangle = \arg\min \int (\Delta A)^2(\Omega) p(\Omega) d\vec{\Omega} \qquad (2)$$

where $\Omega$ is the angle corresponding to direction $\vec{v}$, $\vec{d\Omega} \| \vec{v}$, $p(\Omega)$ is the angular part of $P(\vec{\xi}, \sigma)$. In general, the area $A$ corresponds to a volume $V$ and $\Omega$ to a solid angle. Another possible criterion optimizes the worst case loss over all directions:

$$\vec{r}_1 = \arg \inf \sup \left\{ (\Delta A)^2 \right\}, \tag{3}$$

etc. In what follows the point $\vec{r}_0$ is called the 'Bayes-point'.

The calculation of the Bayes-point scales still exponentially with increasing dimension. Therefore, one must look for estimates of $\vec{r}_0$ which are easier to compute. The proposal of T. Watkin [4] is to consider instead (2) the center of mass of the polygon,

$$\vec{S} = \frac{\int_\mathcal{P} \vec{r} \rho(\vec{r}) dA}{\int_\mathcal{P} \rho(\vec{r}) dA} \tag{4}$$

with the surface mass density $\rho(\vec{r}) = const$. As seen in Table I, this is an excellent *approximation*. However, efficient methods for calculating the center of mass of a polyhedron in high dimensions are lacking. For 'round' polygons the center of the largest inscribed or smallest circumscribed circle should be a good choice. Since $\mathcal{P}$ is represented as a set of inequalities, only the largest inscribed circle is a feasible alternative. This is displayed in Fig. 2 and leads to the estimate contained in Table I. A better approximation is given by the center of the largest volume inscribed ellipsoid [19], a problem also common in nonlinear optimization [20]. The best known algorithms are of order $O(M^{3.5})$ operations, where $M$ is the number of examples and additional logarithmic factors have been neglected (see [20] for details).

The purpose of this paper is to show that a reasonable estimate of $\vec{r}_0$ can be obtained by following the (ergodic) trajectory of an elastic ball *inside* the polygon, as shown in Fig. 3a for four collisions. Fig. 3b shows the trajectory of Fig. 3a from another perspective, by performing an appropriate reflection at each collision edge. A trajectory is periodic if after a finite number of such reflections the polygon $\mathcal{P}$ is mapped onto itself. Fully integrable systems correspond in this respect to polygons which will fill without holes the whole space (that this point of view applies also to other fully integrable systems is nicely exposed in [21]).

If the dynamics is ergodic in the sense discussed in Section 1, then a long enough trajectory should cover without holes the surface of the polygon. By computing the total center of mass of the trajectory one should then obtain a good estimate of the center of mass. The question whether a billiard formed by a 'generic' convex polytope is ergodic or not is to my knowledge not solved. Extensive numerical calculations are possible only in low dimensions. The extent to which the trajectory covers $\mathcal{P}$ after 1000 collisions is visualized in Fig. 4. By continuing this procedure, one can convince oneself that all holes are filled up, so that the trajectory will visit every point inside $\mathcal{P}$.

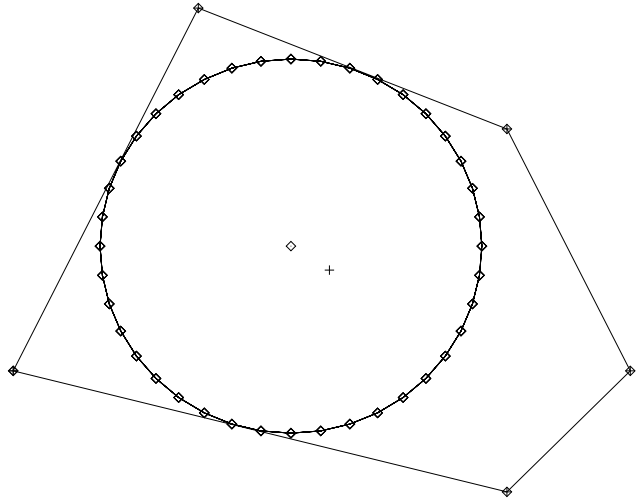

Figure 2: *The largest inscribed circle. The center of mass (cross) is plotted for comparison.*

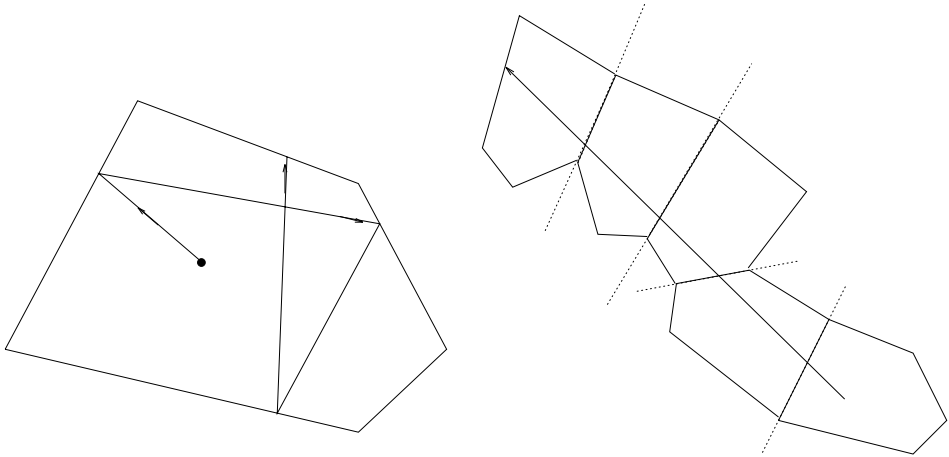

Figure 3: *a) A trajectory with four collisions and b) its 'straightened' form*

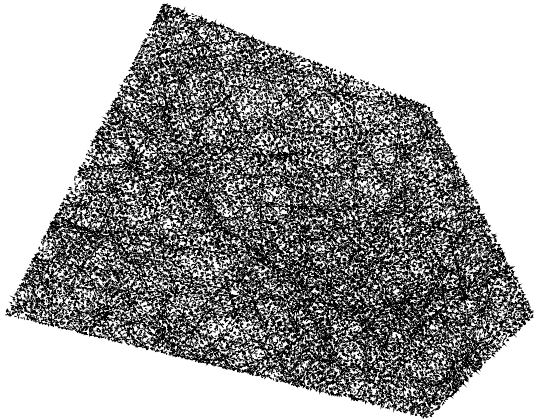

Figure 4: *A trajectory (shown as a dotted line) with 1000 collisions*

The next question is whether the area of $\mathcal{P}$ is *homogeneously* covered by the trajectory. The numerical results summarized in Table I were obtained by averaging over 100 different trajectories and converge to the center of mass as the length of the trajectory is increased.

| Method | x | $\sigma_x$ | y | $\sigma_y$ | $\langle \Delta A^2 \rangle$ |
|---|---|---|---|---|---|
| Bayes-point | 6.1048 | | 1.7376 | | 1.4425 |
| Center of Mass | 6.1250 | | 1.6667 | | 1.5290 |
| Largest inscribed circle | 5.4960 | | 2.0672 | | 7.3551 |
| Billiard - 10 collisions | 6.0012 | 0.701 | 1.6720 | 0.490 | 7.0265 |
| Billiard - $10^2$ collisions | 6.1077 | 0.250 | 1.6640 | 0.095 | 2.6207 |
| Billiard - $10^3$ collisions | 6.1096 | 0.089 | 1.6686 | 0.027 | 1.6774 |
| Billiard - $10^4$ collisions | 6.1232 | 0.028 | 1.6670 | 0.011 | 1.5459 |
| Billiard - $10^5$ collisions | 6.1239 | 0.010 | 1.6663 | 0.004 | 1.5335 |
| Billiard - $10^6$ collisions | 6.1247 | 0.003 | 1.6667 | 0.003 | 1.5295 |

Table I: *Exact value of the Bayes point and various estimates*

## 4 The Perceptron geometry

The Bayes solution of the Perceptron problem requires finding *all* solutions of the inequality set Eq. (1). Consider the convex hulls of the vectors $\vec{\xi}^{(\alpha)}$ belonging to the positive examples $\sigma_\alpha = 1$ and of those in the negative class $\vec{\xi}^{(\beta)}$, $\sigma_\beta = -1$, respectively. The Perceptron with maximal stability can be easily understood by considering the geometric construction shown in Fig. 5.

In addition to the inequalities (1) one has to fulfill the optimality requirement

$$\vec{w} = \arg\max_{\vec{w}} \{\theta_1 - \theta_2\}; \quad (\vec{w}, \vec{w}) = 1 \qquad (5)$$

Geometrically this corresponds to the slab of maximal width one can put between the two convex hulls (the 'maximal dead zone' [22]) and is equivalent (dual) to finding the direction of the shortest line segment connecting the two convex hulls (the minimal connector problem). Choosing $\theta = \frac{\theta_1 - \theta_2}{2}$ (dotted line in Fig. 5 ) one obtains the maximal stability Perceptron (MSP). Notice that the solution is determined by at most $N + 1$ vertices taken from both convex hulls.

A fast algorithm (average of $O(N^2 M)$ operations and storage of order $O(N^2)$) for computing the minimal connector can be found in [8]. In order to switch to version space let us introduce the $N + 1$ dimensional example vectors $\vec{\zeta}^{(\nu)} = (\sigma_\nu \vec{\xi}^{(\nu)}, -\sigma_\nu)$ and the hyperplane vectors $\vec{W} = (\vec{w}, \theta)$. In this

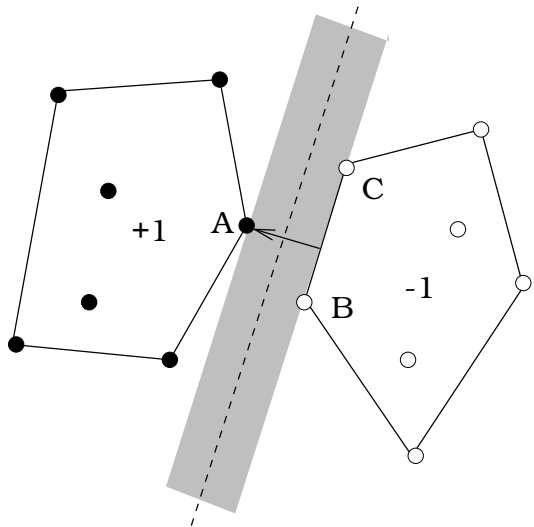

Figure 5: *The Perceptron with maximal stability in example space*

representation Eq. 1 becomes equivalent to the standard set of inequalities

$$(\vec{\mathbf{W}}, \vec{\zeta}^{(\nu)}) > 0 \qquad (6)$$

The version space is the space of vectors $\vec{\mathbf{W}}$. The inequalities (6) define a convex polyhedral cone whose boundary hyperplanes are determined by the training set $\left\{\vec{\zeta}^{(\nu)}\right\}_{\nu=1}^{M}$. Each example constraints in principle the extent of the solution space. More exactly, only the most stringent inequalities form the boundaries of the solution polyhedron, the other examples (hyperplanes) lie outside the polyhedral cone. In order to compactify the solution space one can map the polyhedral cone into the unit-hypersphere by requiring $(\vec{\mathbf{W}}, \vec{\mathbf{W}}) = R^2$.

It is easy to see that the maximal stability Perceptron solution corresponds to inscribing within the solution polyhedron the largest possible sphere. This sphere is tangent to the hyperplanes corresponding to the vertices 'active' in determining the MSP slab (the examples A, B, and C in Fig. 5). The corresponding distance equals the radius $R = \frac{\theta_1 - \theta_2}{2}$, which according to (5) is maximal.

The Bayes decision is taken in version space as following: for each new (test) example generate the corresponding hyperplane in version space. If this hyperplane does not intersect the solution polyhedron, all Perceptrons satisfying Eq. (6) will classify unanimously the new example. If the hyperplane partitions the solution polyhedron in two parts, the decision which minimizes the generalization error is given by evaluating the average measure of *pro* vs the average measure of *contra* votes. Assuming a homogeneous distribution of solutions, we are led to a geometrical problem similar to the one discussed in Section 2. One exception: the solution polyhedral cone is either open or is defined on the unit $N + 1$-dimensional hypersphere.

## 5  How to play billiard in version space

Each billiard game starts by first placing the ball(s) inside the billiard. This is not always a trivial task. In our case, the maximal stability Perceptron algorithm [8] does it or signals that a solution does not exist. The trajectory is initiated by generating a random unit direction $\vec{v}$ in version space.

The basic problem consists of finding out where - on which hyperplane - the next collision will take place. The idea is to compute how much time the ball needs until it eventually hits each one of the $M$ hyperplanes. Given a point $\vec{W} = (\vec{w}, \theta)$ in version space and an unit direction vector $\vec{v}$, let denote the distance along the hyperplane normal $\vec{\zeta}$ by $d_n$ and the component of $\vec{v}$ perpendicular to the hyperplane by $v_n$. Hence, the flight time needed to reach this plane is given by

$$
\begin{aligned}
d_n &= (\vec{W}, \vec{\zeta}) \\
v_n &= (\vec{v}, \vec{\zeta}) \\
\tau &= -\frac{d_n}{v_n}
\end{aligned}
\quad (7)
$$

After computing all $M$ flight times, one looks for the smallest positive $\tau_{min} = \min_{\{\alpha,\beta\}} \tau > 0$. The collision will take place on the corresponding hyperplane. The new point $\vec{W}'$ and the new direction $\vec{v}'$ are calculated as

$$
\vec{W}' = \vec{W} + \tau_{min}\vec{v} \quad (8)
$$
$$
\vec{v}' = \vec{v} - 2v_n\vec{\zeta} \quad (9)
$$

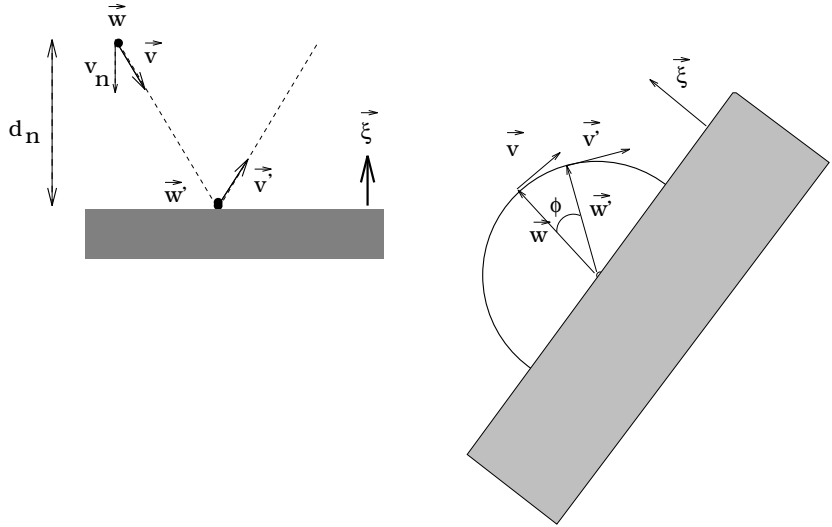

Figure 6: *Bouncing in version space. a) Euclidean b) spherical geometry.*

This procedure is illustrated in Fig. 6a. In order to estimate the center of mass of the trajectory one has first to normalize both $\vec{W}$ and $\vec{W}'$. By using a constant line density one assigns to the (normalized!) center of the segment $\frac{\vec{W}'+\vec{W}}{2}$ the length of the vector $\vec{W}' - \vec{W}$. This is then added to the actual center of mass - as when adding two parallel forces of different lengths. In high dimensions ($N > 5$), however, the difference between the mass of the (normalized) full $N + 1$ dimensional solution polyhedron and the mass of the bounding $N$ dimensional boundaries becomes negligible. Hence, we could as well just record the collision points, assign them the same mass density and constructing their average.

Note that by continuing the trajectory beyond the first collision plane one can sample also regions of the solution space where the $\vec{W}$ makes one, two, etc. mistakes (the number of mistakes equals the number of crossed boundary planes). This additional information can then be used for an optimal decoding when the examples are noisy [3].

Since the polyhedral cone is open, the implementation of this algorithm must take into account the possibility that the trajectory will eventually escape to infinity. The minimal flight time becomes then very large, $\tau > \tau_{max}$.

When this exception is detected a new trajectory is started from the maximal stability Perceptron point in a yet another random direction. Hence, from a practical point of view, the polyhedral solution cone is closed by a spherical shell with radius $\tau_{max}$ acting as a special 'scatterer'. This *flipper* procedure is iterated until gathering enough data.

If we are class conscious and want to remain in the billiard club we must do a bit more. As explained above, the solution polyhedral cone can be closed by normalizing the version space vectors. The billiard is now defined on a curved space. However, the same strategy works also here if between subsequent collisions one follows geodesics instead of straight lines. Fig. 6b illustrates the change in direction for a small time step, leading the the well known geodesic differential equation on the unit sphere:

$$\dot{\vec{W}} = \vec{v} \tag{10}$$
$$\dot{\vec{v}} = - \vec{W} \tag{11}$$

The solution of this equation costs additional resources. Actually, the solution of the differential equation is strictly necessary only when there are no bounding planes on the actual horizon [1]. Once one or more boundaries are 'visible', the choice of the shortest flight time can be evaluated directly, since in the two geometries the flight time is monotonously deformed. In practice, the flipper version is recommended.

Both variants deliver interesting additional informations, like the mean escape time of a trajectory or the number of times a given border plane (example) has been bounced upon. The collision frequency classifies the training examples according to their 'surface' area in the solution polyhedron - a good measure of their relative 'importance'.

## 6 Results and Performance

This Section contains the results of numerical experiments performed in order to test the flipper algorithm. First, the ball is placed inside the billiard with the maximal stability Perceptron algorithm as described in [8]. Next, a number of typically $O(N^2)$ collision points are generated according to the flipper algorithm. Since the computation of one collision point requires $M$ scalar products of $N$ dimensional vectors, the total load of this algorithm is $O(MN^3)$. The choice for $N^2$ collision points is somewhat arbitrary and is based on the following considerations. When using the flipper algorithm we generate many collision points lying on the borders of the solution polyheder. We could try to use this information and approximate the version space polyhedron with an ellipsoidal cone. The number of free parameters involved in the fit is of the order $O(N^2)$. Hence, at least a constant times that many points are needed. The fitted ellipsoid delivers also an estimate on the decision uncertainty. If one is not interested on this information, by monitoring the moving average of some of the trajectory center of mass projections one could stop the billiard after a stable solution is found. For example, Fig. 7 shows the moving average of the generalization probability for a $O(N^2)$ long trajectory. The measurement was made for an input dimension $N = 100$ and $\alpha = \frac{M}{N} = 10$.

---
[1] Assuming light travels along Euclidean straight lines

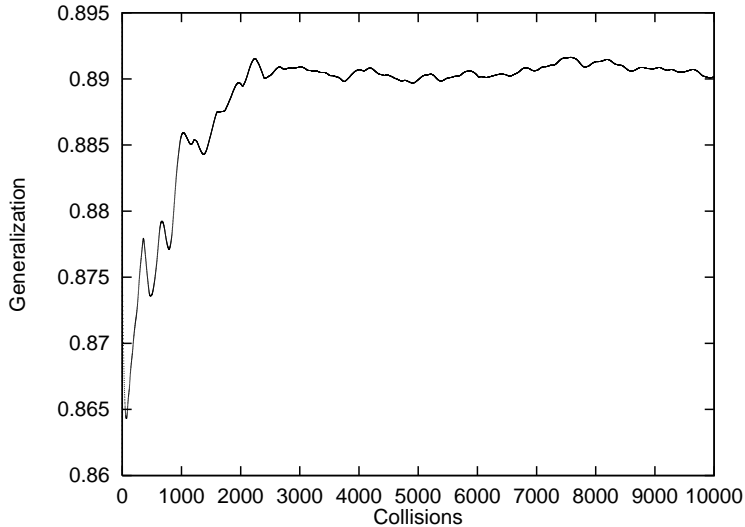

Figure 7: *Moving average of the generalization probability calculated from the trajectory center of mass, $N = 100$, $M = 1000$*

In a first test a known ('teacher') Perceptron was used to label the randomly generated examples for training. The generalization probability was then computed by measuring the overlap between the resulting solution ('stu-dent') Perceptron with the teacher Perceptron. The numerical results obtained from 10 different realizations for $N = 100$ are compared with the theoretical Bayes-learning curve in Fig. 8. Fig. 9 shows a comparison between the billiard results and the maximal stability Perceptron (MSP) results. Although the differences seem small compared to the error bars, the billiard solution was in all realizations consistently superior to the MSP.

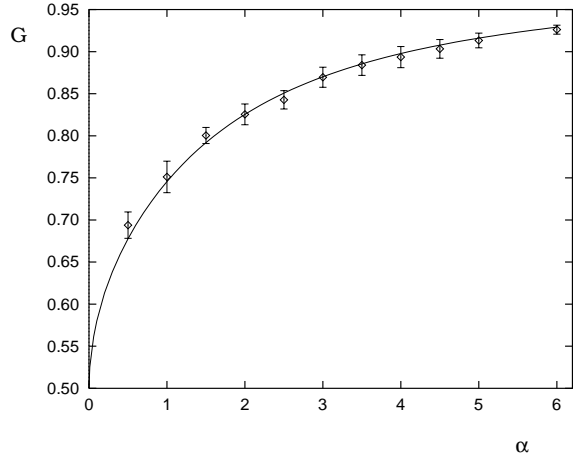

Figure 8: *The theoretical Bayes-learning curve (solid line) vs billiard results obtained in 10 independent trials. $G(\alpha)$ is the generalization probability, $\alpha = \frac{M}{N}$, $N = 100$.*

Fig 10 shows how the number of constraints (examples) on the border of the version polyhedron changes with increasing $\alpha = \frac{M}{N}$.

As the number of examples increases, the probability of escape from the solution polyhedron decreases, the network reaches its storage capacity. Fig. 11 shows the average number of collisions before escape as a function of classification error, parameterized through $\alpha = \frac{M}{N}$. Therefore, by measuring either the escape rate or the number of 'active' examples we can estimate the generalization error *without* using test examples. Note, however, that this kind of calibration graphs might be different for correlated inputs.

Randomly generated training examples sampled from a known teacher Perceptron lead to rather isotropic polyhedra, as illustrated by the small difference between the Bayes- and the MSP learning curve (see Fig. 9). Hence, we expect that the flipper algorithm leads to bigger improvements when applied to real-life problems with strongly anisotropic solution polyhedra. In addition, in many nonlinearly separable problems one can construct multilayer Perceptron networks by using iteratively the Perceptron algorithm [23]. Such procedures have been used, for example, for classifying handwritten digits [24]. For all such examples sets available to me, the introduction of the billiard algorithm on top of the maximal stability Perceptron leads to consistent improvements of up to 5% in classification probability.

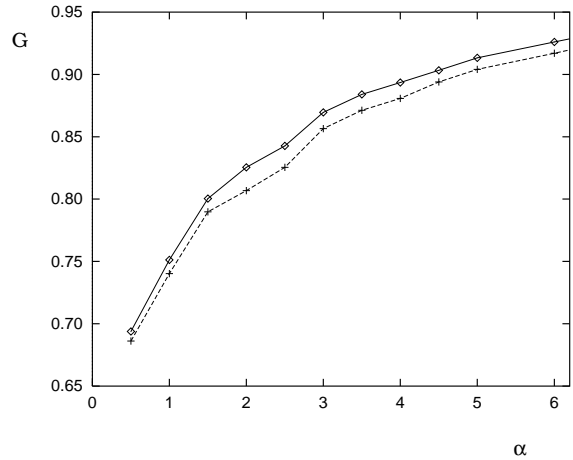

Figure 9: *Average generalization probability, same parameters as above. Lower curve: the maximal stability Perceptron algorithm. Upper curve: the billiard algorithm*

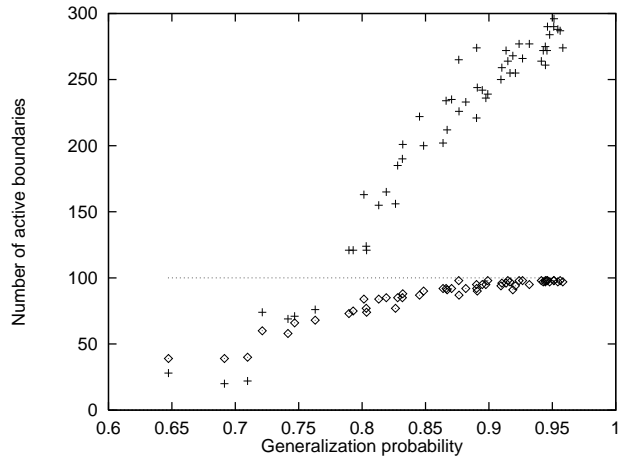

Figure 10: *Number of different identified polyhedral borders vs $G(\alpha)$, $N = 100$. Diamonds: maximal stability Perceptron, saturating at 101. Crosses: billiard algorithm.*

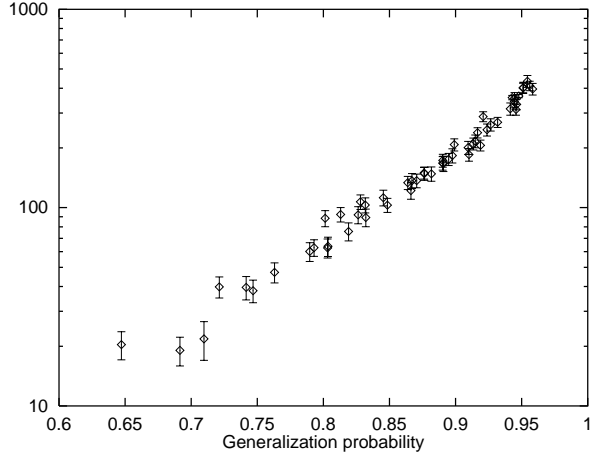

Figure 11: *Mean number of collisions before escaping the solution polyhedral cone vs $G(\alpha)$, $N = 100$*

A publicly available data set known as the sonar-problem [25, 26] considers the problem of deciding between rocks and mines from sonar data. The input space is $N = 60$ dimensional and the whole set consists of 111 mine and 97 rock examples. By using every other data point as training set, and the rest as a test set [25] reports the results contained in Table II for different feedforward architectures. By applying the MSP algorithm we found first that the whole data (training + test) set is linearly separable. Second, by using the MSP on the training set we obtain a 77.5% classification rate on the test set (compared to 73.1% in [25]). Playing billiard leads to a 83.4% classification rate (in both cases the training set was faultlessly classified). This improvement amount is typical also for other applications. The num-ber of active examples (those contributing to the solution) was 42 for the maximal stability Perceptron and 55 during the billiard.

By computing the maximal stability and the Bayes Perceptrons we did not use any information available on the test set. On the contrary, many networks trained by backpropagation are slightly 'adjusted' to the test set by changing network parameters - output unit biases and/or activation function decision bounds. Also other training protocols allow either such adjustments or generate a population of networks, from which a 'best' is chosen based on test set results. Although such adaptive behavior might be advantageous in many practical applications, it is misleading when trying to infer the real capability of the trained network.

In the sonar problem, for instance, we know that a set of Perceptrons separating faultlessly the whole data (training + test) set is included in the version space. Hence, one could use the billiard or other method to find it. This would be an extreme example of 'adapting' our solution to the test set. Such a procedure is especially dangerous when the test set is not 'typical'. Since in the sonar problem the data was divided in two equal sets, by exchanging the roles of the training and test sets one would expect similar quality results. However, we obtain in this case much weaker results (73.3% classification rate). This shows that the two sets do not contain the same amount of information about the common Perceptron solution.

| Hidden Units | % Right on Training set | Std. Dev. | % Right on Test Set | Std. Dev. |
| --- | --- | --- | --- | --- |
| 0 | 79.3 | 3.4 | 73.1 | 4.8 |
| 0–MSP | 100.0 | - | 77.5 | - |
| 0–Billiard | 100.0 | - | 83.4 | - |
| 2 | 96.2 | 2.2 | 85.7 | 6.3 |
| 3 | 98.1 | 1.5 | 87.6 | 3.0 |
| 6 | 99.4 | 0.9 | 89.3 | 2.4 |
| 12 | 99.8 | 0.6 | 90.4 | 1.8 |
| 24 | 100.0 | 0.0 | 89.2 | 1.4 |

Table II. *Results for the sonar classification problem from [23]. 0-MSP is the maximal stability Perceptron, 0-Billiard is the Bayes billiard estimate.*

Looking at the results of Table II it is hard to understand how 105 training examples could substantiate the excellent *average* test set error of 90.4±1.8% for networks with 12 hidden units (841 free parameters). The method of structural risk minimization [6] uses uniform bounds for the generalization error in networks with different VC-dimensions. To firmly establish a classification probability of 90% one needs about ten times more training examples already for the linearly separable class of functions. A network with 12 hidden units has certainly a much larger capacity and requires that many more examples. One could argue that each of the 12 hidden units has solved the problem on its own and thus the network acts as a committee machine. However, such a majority decision should be at best comparable to the Bayes estimate.

# 7 Conclusions and Prospects

The study of dynamic systems led already to many interesting practical applications in time-series analysis, coding, and chaos control. The elementary application of Hamiltonian dynamics presented in this paper demonstrates that the center of mass of a long dynamic trajectory bouncing back and forth between the walls of the convex polyhedral solution cone leads to a good estimate of the Bayes-decision rule for linearly separable problems.

Somewhat similar ideas have been recently applied to constrained nonlinear programming [28].

In fact, the Perceptron problem (1) is similar to the linear inequalities problem and hence has the same algorithmic complexity as the linear programming problem. The theory of convex polyhedra plays a central role both in mathematical programming and in solving $\mathcal{NP}$-hard problems such as the traveling salesman problem [27].

Viewed from this perspective, the ergodic theory of convex polyhedra might provide new, effective tools for solving hard combinatorial optimization problems. Another advantage of such algorithms is that they can be run in parallel in a simple, natural way.

The success of further applications depends, however, on methods of making such simple dynamics strongly mixing. On the theoretical side more general results, applicable to large classes of convex polyhedral billiards are called for. In order to bound the average behavior of 'ergodic' algorithms a good estimate of the average escape (or typical mixing) time is hardly needed.

# Acknowledgments

I thank Manfred Opper for the analytic Bayes data and Bruno Eckhardt for discussions.